%% file: manuscript.tex
\documentclass[10pt,journal,compsoc]{IEEEtran}
\pdfoutput=1
\usepackage{ifpdf}
\ifCLASSOPTIONcompsoc
  \usepackage[nocompress]{cite}
\else
  \usepackage{cite}
\fi
\ifCLASSINFOpdf
  \usepackage[pdftex]{graphicx}
  \graphicspath{{./figs/}}
  \DeclareGraphicsExtensions{.pdf,.jpeg,.jpg,.png}
\else
\fi
\usepackage{amsmath}
\usepackage{algorithm}
\usepackage{algpseudocode}

\usepackage{array}
\ifCLASSOPTIONcompsoc
  \usepackage[caption=false,font=footnotesize,labelfont=sf,textfont=sf]{subfig}
\else
  \usepackage[caption=false,font=footnotesize]{subfig}
\fi
\usepackage[colorinlistoftodos]{todonotes}

\usepackage{enumerate}

\usepackage{rotating}
\usepackage{multirow}

\begin{document}
%
% paper title
% Titles are generally capitalized except for words such as a, an, and, as,
% at, but, by, for, in, nor, of, on, or, the, to and up, which are usually
% not capitalized unless they are the first or last word of the title.
% Linebreaks \\ can be used within to get better formatting as desired.
% Do not put math or special symbols in the title.
\title{Advancing the State-of-the-Art in \\ Hardware Trojans Design}
%
%
% author names and IEEE memberships
% note positions of commas and nonbreaking spaces ( ~ ) LaTeX will not break
% a structure at a ~ so this keeps an author's name from being broken across
% two lines.
% use \thanks{} to gain access to the first footnote area
% a separate \thanks must be used for each paragraph as LaTeX2e's \thanks
% was not built to handle multiple paragraphs
%
%
%\IEEEcompsocitemizethanks is a special \thanks that produces the bulleted
% lists the Computer Society journals use for "first footnote" author
% affiliations. Use \IEEEcompsocthanksitem which works much like \item
% for each affiliation group. When not in compsoc mode,
% \IEEEcompsocitemizethanks becomes like \thanks and
% \IEEEcompsocthanksitem becomes a line break with idention. This
% facilitates dual compilation, although admittedly the differences in the
% desired content of \author between the different types of papers makes a
% one-size-fits-all approach a daunting prospect. For instance, compsoc 
% journal papers have the author affiliations above the "Manuscript
% received ..."  text while in non-compsoc journals this is reversed. Sigh.

\author{
Syed~Kamran~Haider,
~Chenglu~Jin,
and~Marten~van~Dijk% <-this % stops a space
\IEEEcompsocitemizethanks{\IEEEcompsocthanksitem S. K. Haider, C. Jin, and M. van Dijk are with the Department
of Electrical and Computer Engineering, University of Connecticut, Storrs,
CT, 06279.\protect\\
% note need leading \protect in front of \\ to get a newline within \thanks as
% \\ is fragile and will error, could use \hfil\break instead.
E-mail: \{syed.haider, chenglu.jin, vandijk\}@engr.uconn.edu}% <-this % stops an unwanted space
%\thanks{Manuscript received April 19, 2005; revised August 26, 2015.}
}

\IEEEtitleabstractindextext{%
\begin{abstract}
\input{sections/abstract}

\end{abstract}

% Note that keywords are not normally used for peerreview papers.
}

% make the title area
\maketitle

% To allow for easy dual compilation without having to reenter the
% abstract/keywords data, the \IEEEtitleabstractindextext text will
% not be used in maketitle, but will appear (i.e., to be "transported")
% here as \IEEEdisplaynontitleabstractindextext when the compsoc 
% or transmag modes are not selected <OR> if conference mode is selected 
% - because all conference papers position the abstract like regular
% papers do.
\IEEEdisplaynontitleabstractindextext
% \IEEEdisplaynontitleabstractindextext has no effect when using
% compsoc or transmag under a non-conference mode.

% For peer review papers, you can put extra information on the cover
% page as needed:
% \ifCLASSOPTIONpeerreview
% \begin{center} \bfseries EDICS Category: 3-BBND \end{center}
% \fi
%
% For peerreview papers, this IEEEtran command inserts a page break and
% creates the second title. It will be ignored for other modes.
\IEEEpeerreviewmaketitle

\newtheorem{theorem}{Theorem}
\newtheorem{lemma}{Lemma}
\newtheorem{definition}{Definition}

%*******************************************************%
%          INCLUDE SECTIONS                                                           %
%*******************************************************%
\input{sections/introduction}

\input{sections/background}

\input{sections/trojan_class}
\input{sections/evaluation}

\input{sections/related_work}
\input{sections/conclusion}

\ifCLASSOPTIONcaptionsoff
  \newpage
\fi

\bibliographystyle{IEEEtran}
\bibliography{sections/refs} 

\end{document}

%% file: sections/abstract.tex
Electronic Design Automation (EDA) industry heavily reuses third party IP cores.
These IP cores are vulnerable to insertion of Hardware Trojans (HTs) at design time by third party IP core providers or by malicious insiders in the design team.
%These Trojans must be detected in pre-silicon phase, otherwise an adversary can infect millions of ICs through a Trojan affected IP core.
State of the art research has shown that existing HT detection techniques, which claim to detect all publicly available HT benchmarks, can still be defeated by carefully designing new sophisticated HTs.
The reason being that these techniques consider the HT landscape to be limited only to the publicly known HT benchmarks, or other similar (simple) HTs.
However the adversary is not limited to these HTs and may devise new HT design principles to bypass these countermeasures.

In this paper, we discover certain crucial properties of HTs which lead to the definition of an exponentially large class of Deterministic Hardware Trojans $H_D$ that an adversary can (but is not limited to) design.
The discovered properties serve as HT design principles, based on which we design a new HT called \textit{XOR-LFSR} and present it as a `proof-of-concept' example from the class $H_D$.
These design principles help us understand the tremendous ways an adversary has to design a HT, and show that the existing publicly known HT benchmarks are just the tip of the iceberg on this huge landscape.
This work, therefore, stresses that instead of guaranteeing a certain (low) false negative rate for a small \textit{constant} set of publicly known HTs, a rigorous HT detection tool should take into account these newly discovered HT design principles and hence guarantee the detection of an \textit{exponentially large} class (exponential in number of wires in IP core) of HTs with negligible false negative rate.

%% file: sections/introduction.tex
\section{Introduction} \label{sec:intro}

%\IEEEPARstart{M}{odern} 
Modern electronic systems heavily use third party IP (intellectual property) cores as their basic building blocks.
An IP core is a reusable block of logic, tailored to perform a particular operation in an efficient manner, that is an intellectual property of one party.
Optimized for area and performance, the IP cores are essential elements of design reuse in electronic design automation (EDA) industry and save a lot of resources to redesign the components from scratch. 
UARTs, DSP units, Ethernet controllers and PCI interfaces etc. are some examples of the IP cores~\cite{ipcore2}.

The IP cores are typically offered either in a hardware description language (e.g., Verilog or VHDL) as synthesizable RTL (also called `open source' IPs) or as generic gate-level netlists (also called `closed source' IPs).
These IP cores give rise to a critical security problem: how to make sure that the IP core does not contain a \textit{Hardware Trojan} (HT)?
A (compromised) IP core vendor acting as an adversary could implant a malicious circuitry in the IP core for privacy leakage or denial of service attacks.

A significant amount of research has been done during the past decade to design efficient tools for HT detection.
Hicks \textit{et al.}~\cite{sp2010} proposed \textit{Unused Circuit Identification} (UCI) which centers on the fact that the HT circuitry mostly remains inactive within a design, and hence such minimally used logic can be distinguished from the other parts of the circuit.
However later works~\cite{uci_1},~\cite{UCI_sp11} showed how to design HTs which can defeat the UCI detection scheme.
Zhang \textit{et al.}~\cite{veritrust} and Waksman \textit{et al.}~\cite{fanci} proposed detection schemes called VeriTrust and FANCI respectively and showed that they can detect all HTs from the TrustHub~\cite{trust_hub} benchmark suite.
Yet again, the most recent technique called DeTrust~\cite{detrust} introduces new Trojan designs which can evade both VeriTrust and FANCI.

The reason behind this cat-and-mouse game between attackers and defenders is that the current HT detection tools offer critically low HT coverage and typically only cover a small \textit{constant} set of \textit{publicly known} HT benchmarks such as TrustHub.
Whereas an adversary may design new HTs which are different from the publicly known HTs in that they can bypass the detection tool, as demonstrated by DeTrust~\cite{detrust}.
It is unclear to what extent the existing HT detection tools are effective for \textit{publicly unknown} HTs.

%\todo[inline]{Add importance of taxonomies in defining scope}

The first and foremost challenge in designing an effective HT detection technique is to define the scope of the countermeasure on the landscape of HTs.
The HT taxonomies can be used for this purpose.
Bhunia {\em et al.} \cite{ieeesurvey} presents a taxonomy of HTs
%, their general models and a taxonomy of the existing countermeasures 
based on an extensive survey of the existing literature.
The trigger based HTs are subdivided into digital and analog categories where digital HTs have two general models, combinational and sequential HTs.
Combinational HTs are those whose trigger circuitry is simply a comparator whereas sequential HTs have memory elements as well in their trigger circuitry.
Trojans are also classified based on payloads, i.e. digital, analog and others such as denial of service.
%Similarly, HT detection approaches are mainly divided into destructive and non-destructive approaches.
%Logic testing and side-channel analysis are the two major techniques under non-destructive approaches.

Since we are talking about digital IP cores, in this paper we limit ourselves to trigger activated digital HTs which have digital payloads. 
The Trojans that are always active and/or exploit side channels for their payloads are out of scope of this paper.

Even though existing HT taxonomies, such as the one mentioned above, provide significant knowledge about the HT properties, yet this information is quite fundamental and does not provide detailed characteristics vital for the HT countermeasures to provide strong security guarantees.
This limitation may lead to uncertainties and potentially misleading information about the detection coverage that a HT countermeasure can provide.

%This paper introduces of these detailed characteristics of trigger activated digital HTs is the key contribution of this paper which allows the HT detection techniques to provide transparent security guarantees.

%\subsection{Contributions}
This paper introduces four crucial properties ($d$, $t$, $\alpha$, $l$) of a large and complex class $H_D$ of trigger activated and deterministic digital HTs.
These properties, determining the stealthiness of HTs, lead to a much more detailed classification of such HTs and hence assign well defined boundaries to the scope of the existing and new countermeasures on the huge landscape of HTs.
In our model, $H_D$ represents the HTs which are embedded in a digital IP core whose output is a function of only its input, and the algorithmic specification of the IP core can exactly predict the IP core behavior.
A brief highlight of the properties of $H_D$ is as follows:

{
\vspace{6pt}
\renewcommand{\arraystretch}{1.3}
\centering
\begin{tabular}{c p{0.85\columnwidth}} %\hline
$d$: &  \textbf{Trigger Signal Dimension} represents the number of wires used by HT trigger circuitry to activate the payload circuitry in order to exhibit malicious behavior. Large $d$ shows a complicated trigger signal, hence harder to detect. \\ %\hline
$t$: & \textbf{Payload Propagation Delay} is the number of cycles required to propagate malicious behavior to the output port \textit{after} the HT is triggered. Large $t$ means it takes a long time after triggering until the malicious behavior is seen, hence less likely to be detected during testing. \\ %\hline
$\alpha$: & \textbf{Implicit Behavior Factor} represents the probability that given a HT gets triggered, it will not (explicitly) manifest malicious behavior; this behavior is termed as implicit malicious behavior. Higher probability of implicit malicious behavior means higher stealthiness. \\ %\hline
$l$:  & \textbf{Trigger Signal Locality} shows the spread of trigger signal wires of the HT across the IP core. Small $l$ shows that these wires are in the close vicinity of each other. Large $l$ means that these wires are spread out in the circuit, hence the HT is harder to detect. \\ %\hline
\end{tabular}
\vspace{3pt}
}

\begin{figure}[!t]
\centering
\includegraphics[width=0.7\columnwidth]{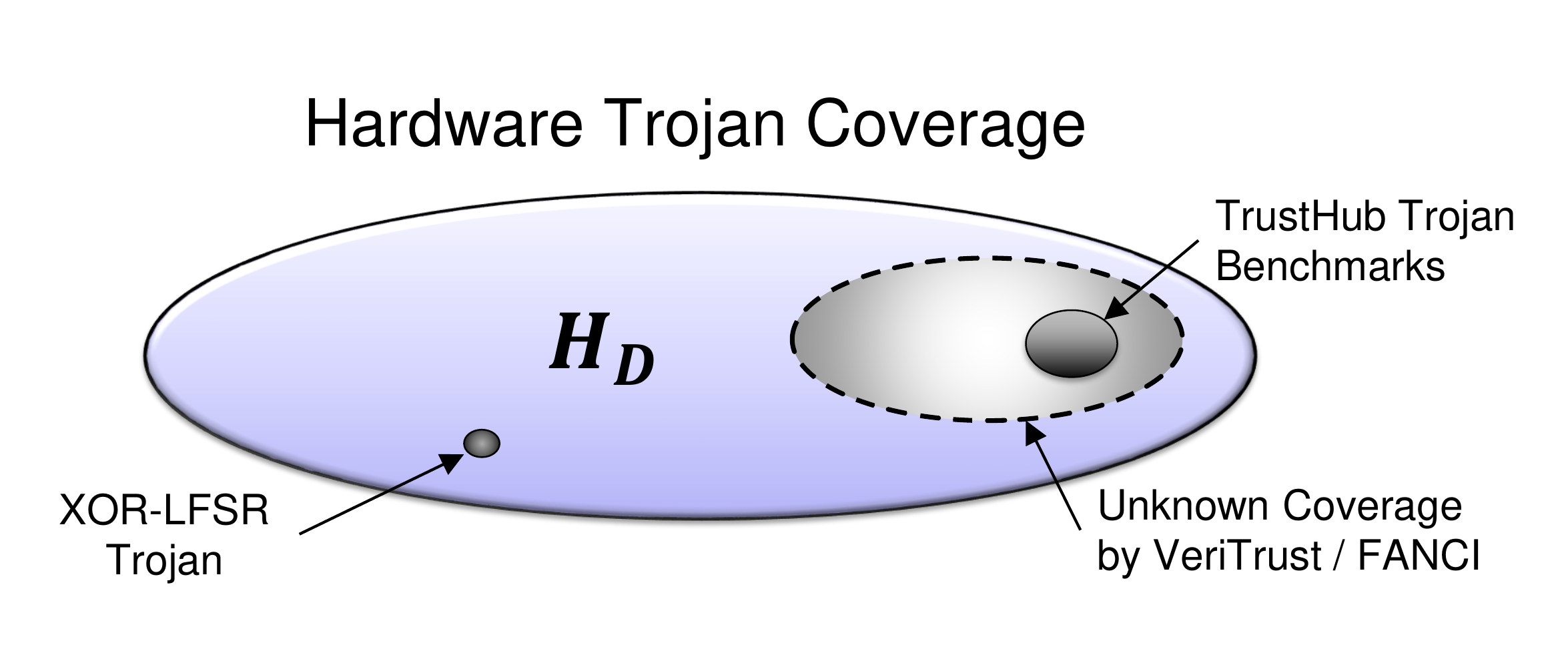}
\caption{Class of Deterministic Hardware Trojans $H_D$, and detection coverages of existing Hardware Trojan countermeasures.}
\label{fig-coverage}
\end{figure}

Based on the above mentioned parameters, we introduce a new stealthy HT, coined the \textit{XOR-LFSR}, which cannot be efficiently detected by ordinary means (design knowledge of the HT itself needs to be incorporated in the detection tool).
%By incorporating the knowledge and understanding of these parameters, HT detection schemes can detect orders of magnitude larger HT subclasses.
%We demonstrate that logic testing based traditional HT detection tools can be bypassed by ``stealthy'' HTs (e.g., those with a large $d$ value) and 
We also show that the current publicly known HTs have small $d$ (mostly $d=1$ for TrustHub) and hence they are the simplest ones.
\figurename~\ref{fig-coverage} depicts a pictorial representation of the HT class $H_D$ that includes TrustHub\footnote{In this paper, we are only considering HTs from TrustHub that are trigger activated and have digital payloads.}.
Although VeriTrust and FANCI can detect TrustHub HTs, it is unclear what security guarantees they offer outside TrustHub.

The rest of this paper is organized as follows.
Section \ref{sec:background-org} provides some basic background of HTs and introduces our threat model.
Section \ref{sec:classification} defines the HT class $H_D$ and introduces several advanced properties of this class which lead to new HT design methodologies.
%Sections \ref{sec:hatch_algo} and \ref{sec:hatch_framework} present the basic algorithm and the detailed implementation of HaTCh respectively.
%In section \ref{sec:comparisons} we compare HaTCh with state of the art HT detection schemes.
%Section \ref{sec:evaluation} shows the experimental evaluation.
%Related work is presented in section \ref{sec:related_work} and we finally conclude in section \ref{sec:conclusion}.

%% file: sections/background.tex
\section{Background \& Definitions}\label{sec:background-org}

\begin{figure}[!t]
\centering
\includegraphics[width=2.6in]{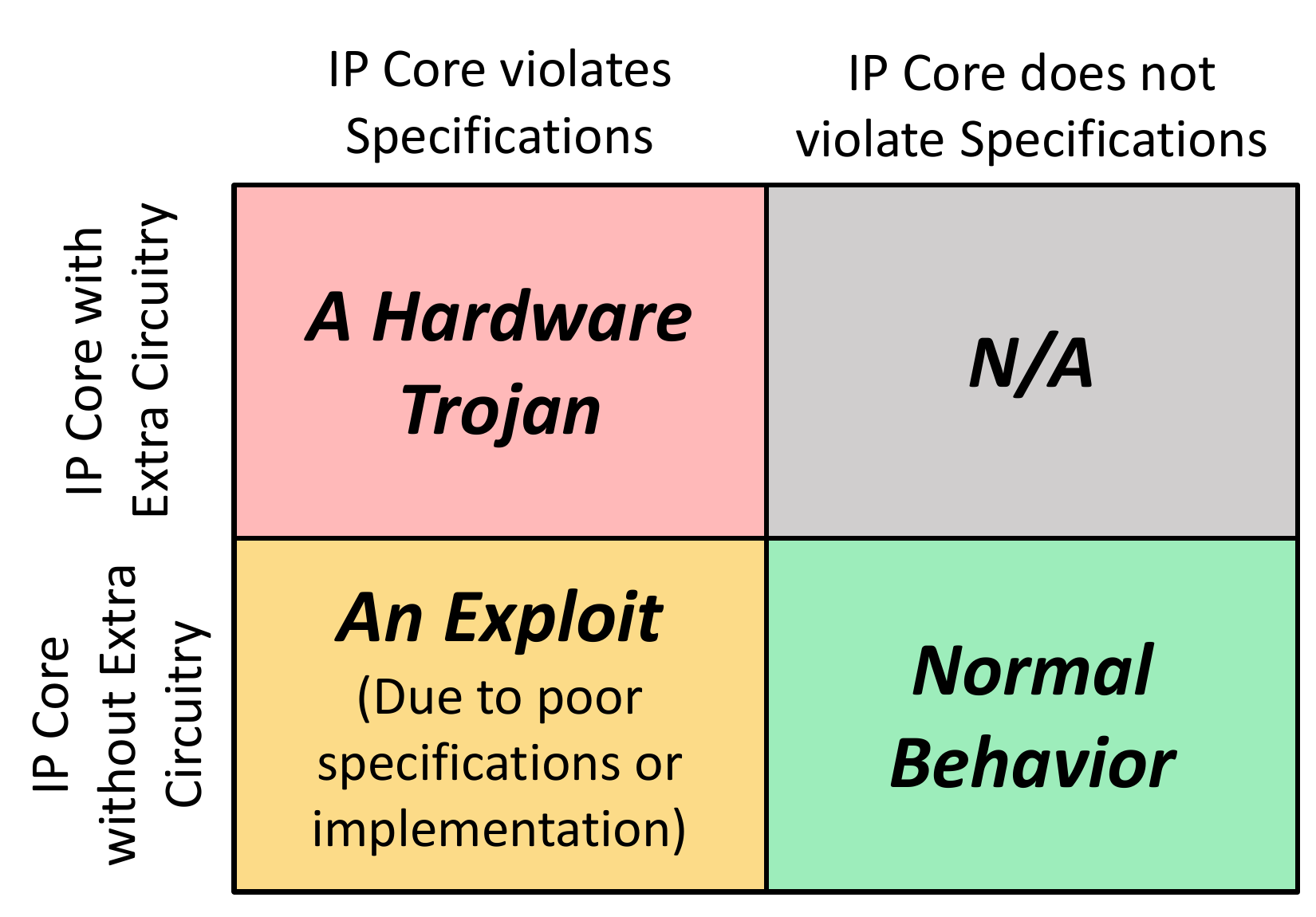}
%\vspace{-6pt}
\caption{IP Core Design Space: An IP core can either have a Hardware Trojan or an Exploit, or it behaves `normally'.}
\label{fig:classification_a}
%\vspace{-12pt}
\end{figure}

%\subsection{IP Core Design Space} \label{sec:ip_design_space}
A digital IP core can fall under one of the following three categories, as shown in \figurename~\ref{fig:classification_a}, based on its level of conformity to the design specifications; (1) containing \textit{A Hardware Trojan}, or (2) containing \textit{An Exploit}, or (3) exhibiting \textit{Normal Behavior}. 

\subsection{Hardware Trojan}
A Hardware Trojan (HT) is malicious \textit{extra} circuitry embedded inside a larger circuit, which results in data leakage or harm to the normal functionality of the circuit once activated.
We define \textit{extra} circuitry as redundant logic added to the IP core without which the core can still meet its design specifications\footnote{Design specifications can also cover the performance requirements of the core, and hence pipeline registers etc. added to the core only for performance reasons can also be considered as `necessary' to meet the design specifications and will not be counted towards `extra' circuitry.}.
A \textit{trigger activated} HT activates upon some special event, whereas an \textit{always active} HT remains active all the time to deliver the intended payload.
%Once activated, a trojan can deliver its payload either through standard I/O channels or through side channels.

Trigger activated HTs typically consist of two parts: a \textit{trigger circuitry} and a \textit{payload circuitry}.
The trigger circuitry is implemented semantically as a comparator which compares the value(s) of certain wires(s) of the circuit with a specified boolean value called \textit{trigger condition}.
The HT trigger circuitry sends its comparator's output to the payload circuitry over certain other wire(s) called the \textit{trigger signal}.
Once the trigger signal is asserted, the payload circuitry performs the malicious operation called `payload' as intended by the adversary.

\begin{definition}
\textbf{Trigger condition} refers to an event, manifested in the form of a particular boolean value of certain internal/external wires of the circuit, which activates the HT trigger circuitry.
\end{definition}
\begin{definition}
\textbf{Trigger signal or Trigger State} refers to a collection of physical wire(s) which the HT trigger circuitry asserts in order to activate the payload circuitry once a trigger condition occurs.
\end{definition}

The trigger signal must not be confused with trigger condition; trigger condition is an event which causes the HT activation, whereas trigger signal is the output of trigger circuitry which tells the payload circuitry to show malicious behavior.

\begin{figure}[!t]
\centering
\subfloat[Half Adder]{\includegraphics[width=1.4in]{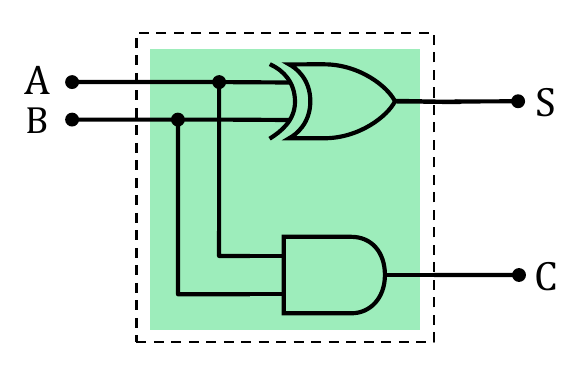}
\label{fig:simple_ht_exp_1}}
\hfil
\subfloat[Half Adder with HT]{\includegraphics[width=1.9in]{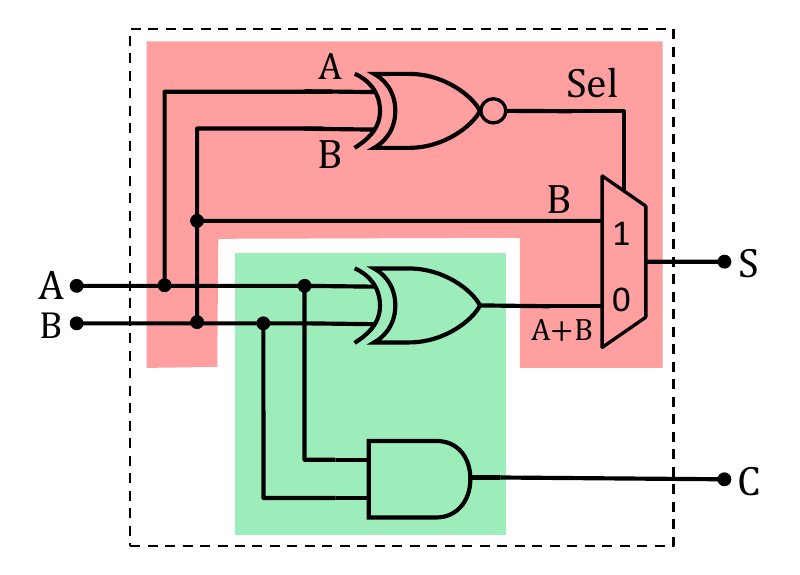}
\label{fig:simple_ht_exp_2}}
\caption{A simple HT: Trigger condition $A=B$; Normal output $S=A \oplus B$; Malicious output (under trigger condition) $S=B$.}
\label{fig:simple_ht_example}
\end{figure}

\figurename~\ref{fig:simple_ht_example} shows an example of a simple HT embedded in a half adder circuit.
The HT-free circuit in \figurename~\ref{fig:simple_ht_exp_1} generates a sum $S=A \oplus B$ and a carry $C=A \cdot B$. 
The HT, highlighted in red in \figurename~\ref{fig:simple_ht_exp_2}, triggers when $A=B$ and produces incorrect results i.e. $S=B$ for $A=B$ and $S=A \oplus B$ for $A \ne B$.
Notice the difference between the trigger condition $A=B$, and the trigger signal $Sel$ which only becomes $1$ when the trigger condition is satisfied.

The Trojan affected circuit in \figurename~\ref{fig:simple_ht_example} produces a \textit{malicious} output $S=1$ for trigger condition $A=B=1$ which is distinguishable from otherwise normal output ($S=0$). 
However, the same circuit produces a (so called) malicious output $S=0$ for trigger condition $A=B=0$ which is the same as otherwise normal output and cannot be distinguished from the `normal' behavior of the circuit.
This observation leads us to the definition of \textit{explicit} vs. \textit{implicit} malicious behaviors:

\begin{definition}
\textbf{Explicit malicious behavior} refers to a behavior of a HT where the HT generated output is \textit{distinguishable} from a normal output.
\end{definition}
\begin{definition}
\textbf{Implicit malicious behavior} refers to a behavior of a HT where the HT generated output is \textit{indistinguishable} from a normal output. 
\end{definition}

Notice that an adversary may exploit the implicit malicious behavior to bypass functional testing based detection tools during the testing phase.
Once the infected circuit has passed the testing and is deployed, it can then manifest explicit malicious behavior to actually deliver the payload.
In other words, implicit malicious behavior can lead to a false negative.

\begin{definition}
\textbf{False Negative} refers to a scenario when a HT detection tool identifies a circuit containing a HT as a Trojan-free circuit or transforms a circuit containing a HT into a circuit which still allows the HT to express implicit or explicit malicious behavior.
\end{definition}

Clearly, all existing dynamic analysis (i.e. functional testing) based approaches which assume that a HT is never triggered during the functional testing phase can suffer from false negatives because of the implicit malicious behavior.
Neglecting the implicit malicious behavior could lead to devastating consequences in a security critical application, e.g. if an adversary designs a HT to significantly increase the probability of implicit malicious behavior and thus alleviates the existing HT detection techniques.

\subsection{An Exploit}
An exploit refers to a loophole in the specifications or implementation of an IP core which allows an adversary to manipulate it beyond its intended specifications.
Notice that an exploitable IP core does not have `extra' circuitry like a Hardware Trojan.
Depending upon the nature of the exploits, they can also deliver the payloads like Hardware Trojans such as privacy leakage or denial of service etc.
Some examples of exploits are as follows: a wireless connection that can be overloaded may lead to a denial of service, a broken AES module due to a predictable key, recent OpenSSL Heartbleed bug etc. 
This paper only focuses on rigorous reasoning about Hardware Trojans (not exploits).

%\subsection{Threat Model}
%A third party IP core is provided in the form of a synthesized netlist which obfuscates the HDL source code.
%The IP core vendor has embedded a trigger based hardware trojan in the IP core which delivers its payload via standard IO channels of the IP core.
%The hardware trojan could have been inserted directly by maliciously modifying the source code or by using some malicious tools to synthesize the RTL.
%This IP core is used in a larger design whose millions of chips are fabricated.
%The adversary wants to exploit this scalability to infect millions of fabricated chips by supplying an infected IP core.
%To prevent this, we test the IP core for potential HTs in \textit{pre-silicon} phase, i.e. before integrating it into a larger design and fabricating it.
%Logic or functional testing, used by Design for Test (DFT) community for testing basic manufacturing defects, is one of the simplest methods to test the IP core for basic HTs which can be easily implemented using the existing simulation/testing tools.
%For the above reasons we restrict ourselves to analyzing logic testing based tools used in pre-silicon phase for HT detection.

%% file: sections/trojan_class.tex
\section{Characterization of Hardware Trojans}\label{sec:classification}

\begin{figure}[!t]
\centering
\includegraphics[width=0.9\columnwidth]{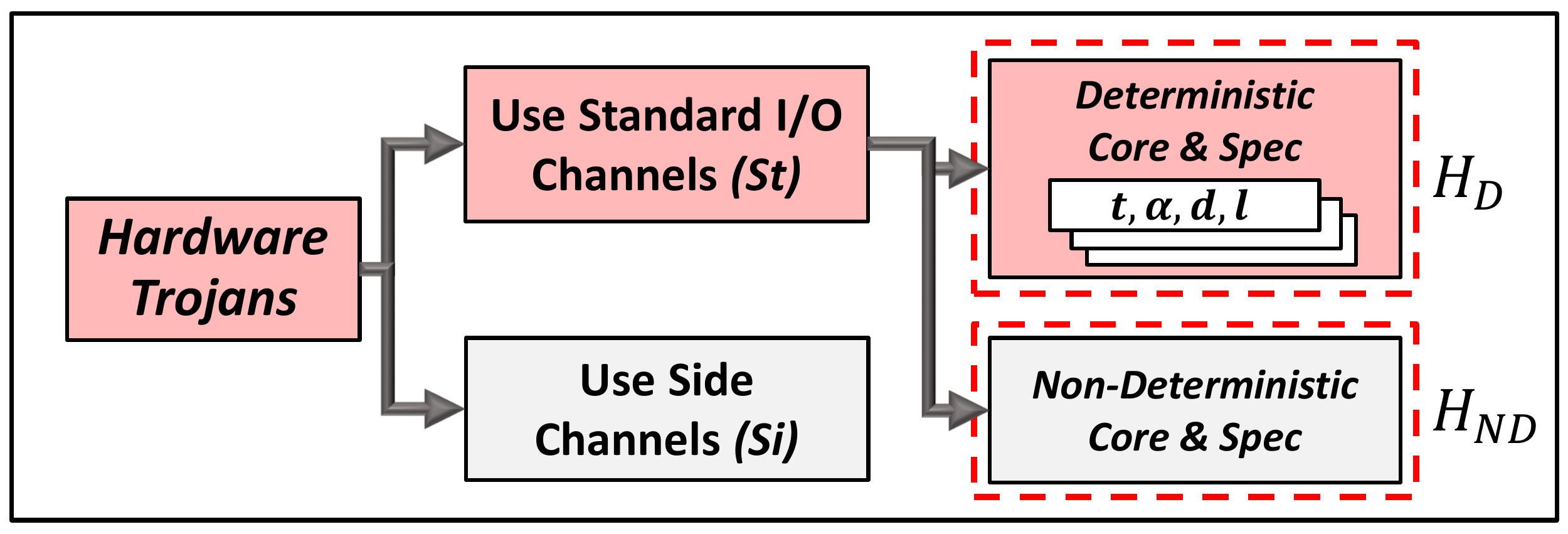} 
%\vspace{-18pt}
\caption{Classification of Hardware Trojans}
\label{fig:taxonomy}
%\vspace{-12pt}
\end{figure}

%Now that we have explained some important terminologies related to HTs which we will be using frequently in the rest of the paper, 
In this section, we characterize the trigger activated HTs based on their following fundamental characteristics that lead to a clear distinction between the deterministic $H_D$ and non-deterministic $H_{ND}$ types.

% ---------------  Standard Channels vs. Side Channels  ----------------------------
\subsection{$St$ vs. $Si$ Hardware Trojans} 
Hardware Trojans are first grouped based on the payload channels they use once activated as shown in  \figurename~\ref{fig:taxonomy}.
\textit{St} refers to the Trojans using only standard I/O channels (this includes LED outputs etc.) whereas \textit{Si} represents the Trojans which also use side channels to deliver the payload.
I/O channels are generally used to communicate binary payloads $b_j$ at certain times $t_j$ for the duration of the execution of the IP core. 
In this sense the view of an I/O channel can be represented as a sequence $(b_1,t_1), (b_2,t_2),\ldots, (b_N,t_N)$. 
Its information is decomposed in three channels: the binary channel corresponding to $(b_1, b_2, \ldots, b_N)$, the timing channel corresponding to $(t_1,t_2,\ldots, t_N)$, and the termination channel $N$ which reveals information about the duration of the execution of the IP core. 
If a Trojan delivers some of its payload over the timing channel (or other side channels), then we define it to be in \textit{Si}. 
If a Trojan delivers {\em all} of its payload using the standard usage of  I/O channels (the binary and termination channels), then we define it to be in \textit{St}.  
E.g., a HT causing performance degradation in terms of slower response/termination times due to slower computation (denial of service in the most extreme case) is in \textit{St}.

% ----------------- Deterministic vs. Non-deterministic IP Core / Specifications ----------------------
\subsection{$H_D$ vs. $H_{ND}$ Hardware Trojans} 
We further refine our description of \textit{St} Trojans by subdividing them in $H_D$ and $H_{ND}$ groups based on the IP core behavior in which they are embedded and their algorithmic specifications.

\begin{definition}
\boldmath{$H_D$} Trojans are the ones which are:
\begin{enumerate} 
\item Embedded in an IP core whose output is a function of only its input -- i.e. the logical functionality of the IP core is deterministic, and 
\item The algorithmic specification of the IP core can exactly predict the IP core behavior.
\end{enumerate}
\end{definition}

If any of the two above mentioned conditions is not satisfied for a \textit{St} type HT then we consider the HT to be in $H_{ND}$.
A true random number generator (TRNG), for example, is a non-deterministic IP core since its output cannot be predicted and verified by logic testing against an expected output.\footnote{Any IP core which contains a TRNG as a module, yet the I/O behavior of the core can still be predicted is considered $H_D$.}
Any \textit{St} Trojan in such a core is considered $H_{ND}$.
A pseudo random number generator (PRNG), on the other hand, is considered a deterministic IP core as its output depends upon the initial seed and is therefore predictable by a logic based testing tool (hence $H_D$).
Similarly, if the algorithmic specification allows coin flips generated by a TRNG then we consider the Trojan to be $H_{ND}$.
On the other hand, if the coin flips are generated by a PRNG then we regard the Trojan as $H_D$.

The non-deterministic behavior of IP cores and/or their functional specification which accepts small probabilistic fluctuations within some acceptable range allows a covert channel for $H_{ND}$ Trojans to embed some minimal malicious payload in the standard output without being detected by an external observer~\cite{bellare2014security}.
The external observer considers these small fluctuations as part of the functional specification.
Hence, the non-deterministic nature of $H_{ND}$ Trojans prohibits the development of a logic testing based tool to detect these Trojans with overwhelming probability.

\section{Advanced Properties of Class $H_D$} \label{sec:HD}

In the following discussion, we introduce some crucial properties of $H_D$ Trojans that characterize their complexity and stealthiness.

\subsection{Trigger Signal Dimension $d$}
When a trigger condition of a hardware Trojan occurs, {\em regardless of the other subsequent user interactions}, its trigger circuitry gets activated and outputs a certain binary value on a certain trigger signal $Trig$ to activate the payload circuitry which manifests malicious behavior.
A trigger signal $Trig$ is represented as a labeled binary vector of one or more wires/registers/flip-flops (each carrying a $0$ or $1$), e.g. in the example circuit of \figurename~\ref{fig:simple_ht_example}, $Sel=1$ is a trigger signal.
In other words, $Trig$ represents a trigger state of the circuit through which the circuit must have passed before manifesting malicious behavior.
Notice that the trigger state(s) can be associated back to the occurrence of the respective trigger condition(s).
Hence a HT can be represented by a set of trigger states ${\cal T}$; i.e. the states which always lead to malicious behavior (implicit or explicit).
\begin{definition} 
\textbf{A set \boldmath{${\cal T}$} of trigger states} \textit{represents} a HT if the HT always passes through one of the states in ${\cal T}$ in order to express implicit of explicit malicious behavior.
\end{definition}

\vspace{3pt}
Notice that {${\cal T}$} is not unique. 
For example, {${\cal T}$} could be {${\cal T} = \{(Sel=1)\}$}, or if $Sel$ is somehow independent of $A$ and $B$, then it could also be {${\cal T}=\{[(A,B)=(1,1)], [(A,B)=(0,0)]\}$} etc.

We define the \textit{trigger signal dimension} $d$ of a HT represented by a set of trigger states ${\cal T}$ as follows:
\begin{definition} 
\textbf{Trigger Signal Dimension \boldmath{$d({\cal T})$}} of a HT is defined as $d({\cal T}) = \max_{Trig \in {\cal T}} |Trig|$.
\end{definition}

\vspace{3pt}
In other words, the trigger signal dimension shows the width of the widest trigger signal bit-vector of a HT.
E.g. for the HT from \figurename~\ref{fig:simple_ht_exp_2}, $d({\cal T})=1$ since the set of trigger states {${\cal T} = \{(Sel=1)\}$} only has the signal $Sel$ that activates the payload circuitry is only 1-bit wide (and hence easy to detect).
In the next section, we show that one can design HTs with high dimensional trigger signals.
%A set of trigger states ${\cal T}$ {\em represents} the hardware Trojan if any of its malicious behaviors must have (at some clock cycle) passed through a state in ${\cal T}$. 
%The dimension of a set of trigger states ${\cal T}$ is defined as $d({\cal T}) =  \max_{Trig \in {\cal T}} |Trig|$. % and 
%Let ${\bf T}$ denote the collection of all possible sets ${\cal 
 % i.e. $Trig\in \{0,1\}^d$, and we call $d$ 
 %  the dimension of a Trojan is defined as 
%  the maximum width of a trigger signal , i.e.
%   $d=\min_{{\cal T}} w({\cal T}) $ where the minimization is over all possible sets ${\cal T}$ of trigger signals that represent the Trojan.
   %(which we also call trigger states) through which malicious behavior must transition.
Obviously, it becomes difficult to detect HTs which only have high dimensional sets of trigger states, i.e. large trigger signals.
The set of possible values of a given trigger signal $Trig$ grows exponentially in $d=|Trig|$ and only one value out of this set can be related to the occurrence of the corresponding trigger condition. 
Clearly, since in theory $d$ can be as large as the number of wires $n$ in the IP core, $H_D$ represents an exponentially (in $n$) large class of possible HTs.

\subsection{Payload Propagation Delay $t$}
For a set ${\cal T}$ which represents a HT, we know that if the HT manifests malicious behavior, then it must have transitioned through a trigger state $Trig\in {\cal T}$ at some previous clock cycle. 
Therefore, we define the \textit{payload propagation delay} $t$ as follows:
\begin{definition}
\textbf{Payload Propagation Delay \boldmath{$t({\cal T})$}} of a hardware Trojan represented by a set of trigger states ${\cal T}$ is defined as the {\em maximum} number of clock cycles taken to propagate the malicious behavior to the output \textit{after} entering a trigger state in ${\cal T}$. 
\end{definition}
%A $H_D$ Trojan that manifests malicious behavior % (consider the explicit malicious behavior for now) 
%must have first seen its \textit{trigger condition} occurring and transitioned through a trigger state $Trig$ from a set of trigger signals ${\cal T}$ which  represents the Trojan. We define  ....... [TO DO]
% of ``no return'' (which we have called a trigger state). 
%Clearly, since the number of possible states is finite, there must be some upper bound $t$ on the number of cycles within which malicious behavior manifests after a ``last trigger state''.
%In other words, $t$ represents the number of clock cycles taken to propagate the malicious behavior of a Trojan to the output port \textit{after} a trigger condition (corresponding to the last state  has occurred, i.e. from the moment when a trigger condition occurs till the moment when its resulting malicious behavior shows up at  the output port.
%For simplicity, we considered only explicit malicious behavior so far; however, the same logic applies to the Trojans with implicit malicious behavior as well.
%Such Trojans also see a trigger condition and then propagate some (malicious) result to the output; although this result happens to be the same as `normal' or `expected' result.
%The definition of $t$ can be further clarified with the following example.

I.e., the number of clock cycles from the moment when a trigger signal is asserted till its resulting malicious behavior shows up at the output port.
E.g., consider a counter-based HT where malicious behavior immediately (during the same clock cycle) appears at the output as soon as a counter reaches a specific value. 
Then, $t(\{Trig\})=0$ for the trigger signal $Trig$ which represents the occurrence of the specific counter value (i.e. trigger condition).
However, notice that any counter value $j$ clock cycles before reaching the `specific value' can also be considered as a trigger signal $Trig$ with $t(\{Trig\})=j$, because eventually after $j$ cycles this $Trig$ manifests the malicious behavior.
%On the other hand if we consider the trigger signal $Trig$ which represents the counter value $j$ clock cycles before reaching the specific value, then $t(\{Trig\})=i$.
%
%trigger condition is that the counter reaches a specific value. 
%In this case, $t$ is the latency between the moment when the counter reaches that specific value (the trigger condition) and the moment when its resulting malicious behavior appears at the output. 
%Note that $t$ is independent of the cycles required by the counter to reach the trigger condition and $t$ is 

To detect a HT with a large value of $t({\cal T })$, the memory requirement and complexity of logic testing based detection tool is increased.
However, for any HT, typically there exists a set of trigger signals which represents the HT and which has a small $t$ because of a small number of register(s) between the trigger signal and the output port.

%For the $k$\textit{-XOR-LFSR} Trojan in \figurename~\ref{fig-htd}, $t=0$ since the output is changed in the same clock cycle in which the Trojan gets triggered.

\subsection{Implicit Behavior Factor $\alpha$}
In addition to previously discussed properties of Trojans, the IP core in which the Trojan is embedded plays a critical role in its stealthiness.
According to the definition of implicit malicious behavior, it may not always be possible to distinguish a malicious output from a normal output just by monitoring the output ports.
Consequently, the implicit malicious behavior adds to the stealthiness of the HT since it creates a possibility of having a false negative under logic testing based techniques.
%For example, consider a Trojan being a malicious 2-to-1 MUX, with one of its inputs connected to a malicious wire and the other one to a normal wire (as  in \figurename~\ref{fig:simple_ht_example}). 
%Suppose  the Trojan  triggers and selects as a result the malicious wire as the MUX output instead of the normal wire. If the normal and malicious input have the same value, then the`malicious' output value is indistinguishable from the normal output value.
%Similarly, hardware Trojans embedded in non-deterministic IP cores can also produce such malicious outputs which seem to be legitimate because of inherent randomness of such IP cores.
We quantify this possibility by defining the \textit{implicit behavior factor} $\alpha$ as follows:
%\begin{definition}
%\label{def:aplha}
%\textbf{Implicit Behavior Factor \boldmath{$\alpha({\cal T})$}} of a HT represented by the set of trigger states ${\cal T}$ is defined as 
%$\alpha({\cal T}) = \underset{Trig\in {\cal T}}{\max} \alpha(Trig)$ where $\alpha(Trig)$ shows the probability that, given the trigger state $Trig$ occurs, it will lead to {\em implicit} malicious behavior.
%\end{definition}
\begin{definition}
\label{def:aplha}
\textbf{Implicit Behavior Factor \boldmath{$\alpha({\cal T})$}} of a HT is the probability that given the HT is triggered, it will manifest {\em implicit} malicious behavior.
\end{definition}
%The effect of the IP core on the stealthiness of a hardware Trojan (which is embedded into it) against logic testing techniques can be represented by the stealthiness factor $\alpha$: For a trigger state $Trig$ we define $\alpha(Trig)$ as the probability given that $Trig$ occurs, this will lead to  {\em implicit} malicious behavior.
%We define $\alpha({\cal T})$ as the {\em maximum} of $\alpha(Trig)$ over $Trig\in {\cal T}$.
% probability that given a trigger state in ${\cal T}$, the Trojan manifests an implicit malicious behavior (which can go undetected by the functional testing), i.e.
%\begin{equation*} \alpha = Prob(Implicit Malicious Behavior | Trigger Occurs) \end{equation*}

In other words, the higher the value of $\alpha$, the lower the chance of detection by logic testing even if the HT gets triggered and hence the higher the overall stealthiness of the HT.
E.g. for the HT from \figurename~\ref{fig:simple_ht_exp_2} with the trigger condition $A=B$, if $(A,B)=(1,1)$ then the malicious output $S=1$ is distinguishable from the normal output (i.e. $S=0$).
However for $(A,B)=(0,0)$, the malicious\footnote{The output is considered malicious because it is generated by an activated hardware Trojan.} output $S=0$ is indistinguishable from the normal output (i.e. $S=0$), i.e. the implicit malicious behavior comes into the picture.
Hence, given that this particular HT activates, the probability that the HT-generated (malicious) output is indistinguishable from the normal output is $\alpha({\cal T}) = 0.5$ which represents the implicit behavior factor of this HT.

%\subsubsection{Notation used for $H_D$ Trojans: $H_{t,\alpha,d}$}
%\begin{figure}[!t]
%\centering
%\includegraphics[width=\columnwidth]{achievable}
%\caption{Effects of changing the parameters $t$ and $d$ on $\alpha$.}
%\label{fig:alpha_tradeoffs}
%\end{figure}

%****************** Locality 
\subsection{Trigger Signal Locality $l$}

\begin{figure}[!t]
\centering
\subfloat[A simple circuit]{\includegraphics[width=2in]{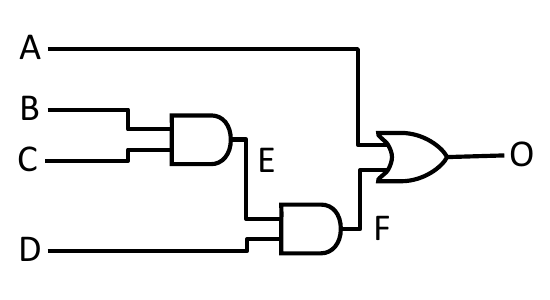}
\label{fig:locality_circuit}}
\hfil
\subfloat[Locality Graph of the circuit]{\includegraphics[width=2in]{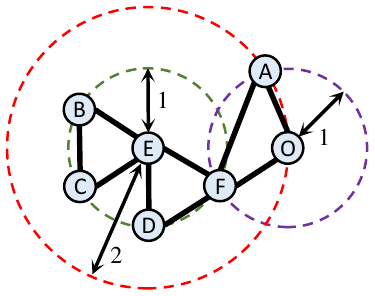}
\label{fig:locality_graph}}
\caption{Example of Locality Graph}
\label{fig:locality_graph_example}
\end{figure}

We notice that the individual wires of a HT trigger signal of dimension $d$ are logically/physically located in the close vicinity of each other in the circuit netlist/layout.
This is because eventually these wires need to coordinate (through some combinational logic) with each other to perform the malicious operation.
Based on this observation, we introduce the idea of \textit{locality} in gate level circuits, similar to the region based approach in \cite{trustedRTL}.

Consider the simple combinational circuit from \figurename~\ref{fig:locality_circuit}.
Based on this circuit, we draw a \textit{locality graph} shown in \figurename~\ref{fig:locality_graph} whose nodes represent the wires of the circuit and each edge between any two nodes represents connectivity of the corresponding two wires through a combinational logic level. 
In other words, each logic gate of the circuit is replaced by multiple edges (three in this case) in the graph which connect together the nodes corresponding to its inputs and the output.
For any two nodes (i.e. wires) $i$ and $j$ in a locality graph, we define $dist(i, j)$ as the shortest distance between $i$ and $j$.
In other words, $dist(i, j)$ represents the minimum number of basic combinational or sequential logic levels (e.g. logic gates and/or flip flops) between wires $i$ and $j$.
E.g. $dist(E, B) = dist(E, C) = 1$, whereas $dist(E, O) = dist(E, A) = 2$.

\begin{definition} \label{def:locality}
\textbf{Trigger Signal Locality \boldmath{$l({\cal T})$}} of a HT represented by the set of trigger states ${\cal T}$ is defined as: 
$$l({\cal T}) = \underset{Trig\in {\cal T}}{\max} \left( \underset{0 \leq i,j < |Trig|}{\max} dist(Trig[i], Trig[j]) \right)$$
where $Trig[i]$ represents the label of the $i_{th}$ wire in $Trig$.
\end{definition}

A low value of $l({\cal T})$ shows that the trigger signal wires of the HT are in the close vicinity of each other and vice versa.
Having a notion of locality can significantly reduce the computational complexity of logic testing based HT detection tools. 

%****************** Achievable Triples  
\subsection{Achievable Quadruples $(d, t,\alpha, l)$}
A Hardware Trojan can be represented by multiple sets of trigger states ${\cal T}$, each having their own $d$, $t$, $\alpha$, and $l$ values.  
The collection of corresponding quadruples $(d,t,\alpha,l)$ is defined as the achievable region of the Hardware Trojan.
%We denote by $H_{d,t,\alpha,l}$ all  $H_D$ type Trojans which can be represented by a set of trigger states ${\cal T}$ with parameters $d({\cal T})\leq d$, $t({\cal T})\leq t$, $\alpha({\cal T})\leq \alpha$ and $l({\cal T})\leq l$.
%In the remainder of this paper we develop HaTCh which takes parameters $t$, $\alpha$ and $d$ as input in order to detect hardware trojans from $H_{t,\alpha,d}$. 
%E.g., by taking $t=0$ we can detect a simple counter-based HT for small $d$ (as we have seen there exists a trigger state $Trig$ for $t=0$ in a simple counter-based HT; HaTCh for $t=0$ will characterize $Trig$ so that malicious behavior can be detected). 
%However, for our more complex $k$\textit{-XOR-LFSR} trojan, HaTCh for $t=0$ only detects this trojan if $d$ is taken $\geq \log k -2\log\log k$.
% which may make HaTCh's computational complexity prohibitive (see Theorem \ref{th:complexity}).

%\subsubsection{Effective Stealthiness: \boldmath{$\alpha^*$}}

The choice of parameters $d$ and $t$ significantly affects $\alpha$ of the HT. % and consequently the detection capability of HaTCh.
%Although reducing $t$ and/or $d$ reduces the memory requirements of the tool and/or computational complexity respectively, it also increases the chances of not detecting the trojan.
%In other words, this can potentially increase the stealthiness factor $\alpha$ of a $H_{t,\alpha,d}$ trojan to $\alpha^* \ge \alpha$.
%We call the probability $\alpha^*$ `effective stealthiness' of the $H_{t,\alpha,d}$ trojan as seen by \Call{HaTCh}{} algorithm as a result of the provided values of parameters $t$ and $d$.
%\figurename~\ref{fig:alpha_tradeoffs} shows the effects of changing $t$ and $d$ on the minimum possible $\alpha$ in the achievable region of a hardware trojan.
$\alpha$ as a function of $t$ and $d$ is decreasing in both $t$ and $d$.
Reducing $t$ means that explicit malicious behavior may not have had the chance to occur, hence, the probability $\alpha$  that no explicit malicious behavior is seen increases. 
Similarly, reducing $d$ can increase $\alpha$ since as a result of smaller $d$, there may not exist a set of trigger signals ${\cal T}$ that represents the HT and satisfies $d({\cal T})\leq d$.
%On the other hand, a higher $d$ incorporates a larger search space and may reduce $\alpha$ down to zero provided that the trojan does not get triggered and manifest implicit malicious behavior during the learning phase.
Increasing $t$ or $d$ only decreases $\alpha$ down to a certain level; the remaining component of $\alpha$ represents the inherent implicit malicious behavior of the HT.

\section{Examples of $H_D$ Trojans}
In this section, we present some examples of new HTs from the class $H_D$.
These HTs demonstrate how the properties of $H_D$ Trojans defined in the previous section play a significant role in determining the stealthiness of these HTs.

\subsection{A Counter-Based $H_D$ Trojan}\label{htd-h2}
\begin{figure}[!t]
\centering
\subfloat[A counter based $H_D$ Trojan. Trigger Condition: $LFSR = 1101$; Trigger Signal: $(W1, W2)$; Payload: Leak $Secret$]{\includegraphics[width=2.4in]{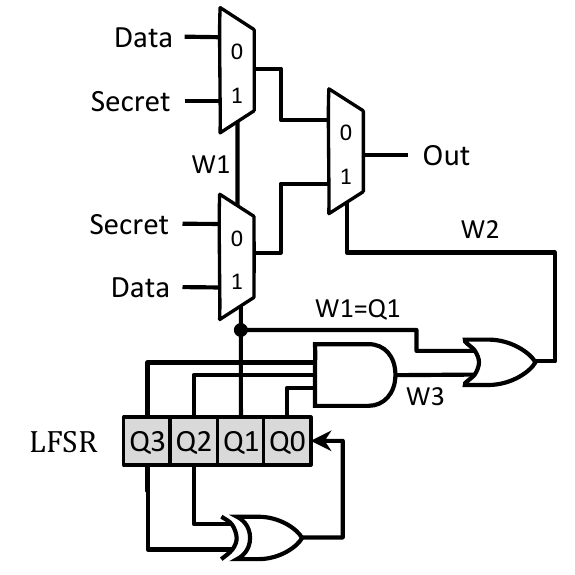}
\label{fig:h2_lfsr_fig}}
\hfill
\subfloat[Truth Table of the Trojan affected circuit. Trigger Condition: $LFSR = 1101$]{
\renewcommand{\arraystretch}{1.1}
\begin{tabular}[b]{|c||c|c|c|c||c|c|c|}
\hline
$Cycle$ & $Q3$ & $Q2$ & $Q1$ & $Q0$ & $W1$ & $W2$ & $W3$ \\
\hline
0 & 1 & 0 & 1 & 0 & 1 & 1 & 0\\
1 & 0 & 1 & 0 & 1 & 0 & 0 & 0\\
2 & 1 & 0 & 1 & 1 & 1 & 1 & 0\\
3 & 0 & 1 & 1 & 1 & 1 & 1 & 0\\
4 & 1 & 1 & 1 & 1 & 1 & 1 & 1\\
5 & 1 & 1 & 1 & 0 & 1 & 1 & 0\\
6 & 1 & 1 & 0 & 0 & 0 & 0 & 0\\
7 & 1 & 0 & 0 & 0 & 0 & 0 & 0\\
8 & 0 & 0 & 0 & 1 & 0 & 0 & 0\\
9 & 0 & 0 & 1 & 0 & 1 & 1 & 0\\
10 & 0 & 1 & 0 & 0 & 0 & 0 & 0\\
11 & 1 & 0 & 0 & 1 & 0 & 0 & 0\\
12 & 0 & 0 & 1 & 1 & 1 & 1 & 0\\
13 & 0 & 1 & 1 & 0 & 1 & 1 & 0\\
\bfseries 14 & \bfseries 1 & \bfseries 1 & \bfseries 0 & \bfseries 1 & \bfseries 0 & \bfseries 1 &\bfseries 1 \\
\hline
\end{tabular}
\label{fig:h2_lfsr_table}
}
\caption{A Counter-Based $H_D$ Trojan having Trigger Signal Dimension $d=2$.}
\label{fig:h2_lfsr_ex}
\end{figure}

The example Trojan shown in \figurename~\ref{fig:h2_lfsr_fig} can leak $Secret$ via $Out$ port instead of $Data$ once it is activated.
The trigger condition of this Trojan is generated by a counter, when reached to ($1101$), which is implemented as a 4-bit maximal LFSR in order to have maximum possible time before the Trojan gets triggered.
The LFSR is initialized to $(Q3,Q2,Q1,Q0)=(1010)$.
At $14th$ clock cycle, the trigger condition (i.e. $LFSR = 1101$) occurs, the HT gets triggered and produces $W1 \ne W2$ on the trigger signal ($W1, W2$).
This results in the activation of payload circuitry (the group of three MUXes) which leaks the secret.

It can be seen in \figurename~\ref{fig:h2_lfsr_table} that all individual wires related to the HT circuitry are continuously showing transitions \textit{without} activating the HT during several clock cycles before reaching to the trigger condition (i.e. cycle $14$).
Therefore, clearly those existing logic testing based HT detection techniques which only look for simplistic one-dimensional HT (i.e. $d=1$) trigger signals do not see any `suspicious' wire which is stuck at $0$ or $1$, and hence this HT is not detected unless it gets activated in a long testing phase. 
This shows that this HT has a trigger signal dimension $d=2$ since two wires $W1$ and $W2$ together constitute the trigger signal.
Since counter based HTs can have large counters, it may not always be feasible to activate them during testing.
Consequently, in order to efficiently detect such HTs, the knowledge of such design parameters must be incorporated while designing the HT countermeasures.

\subsection{An Advanced $H_D$ Trojan: $k$-\textit{XOR-LFSR}}
\begin{figure}[!t]
\centering
\includegraphics[width=2.6in]{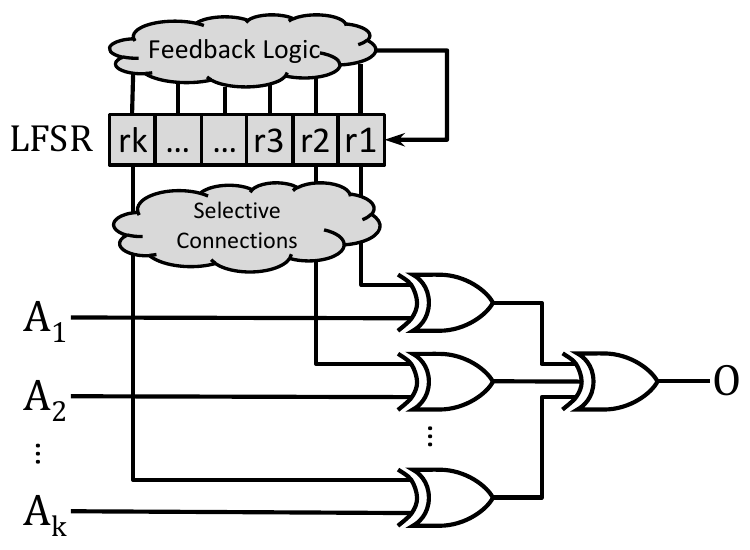}
\caption{$k$-\textit{XOR-LFSR}: A general $H_D$ Trojan.}
\label{fig-htd}
\end{figure}

\figurename~\ref{fig-htd} depicts $k$-\textit{XOR-LFSR}, a counter based Trojan with the counter implemented as an LFSR 
%(with a primitive feedback polynomial) 
of size $k$. The Trojan is merged with the circuitry of an IP core which outputs the XOR of $k$ inputs $A_j$.

Let ${\bf r}^i\in \{0,1\}^k$  denote the LFSR register content at clock cycle $i$ represented as a binary vector of length $k$. 
Suppose that $u$ is the maximum index for which the linear space $L$ generated by vectors ${\bf r}^0, \ldots, {\bf r}^{u-1}$ (modulo 2) has dimension $k-1$. 
Since  $dim(L)=k-1<k=dim(\{0,1\}^k)$, there exists a vector ${\bf v}\in \{0,1\}^k$ such that, (1) the inner products $\langle {\bf v},{\bf r}^i\rangle=0$ (modulo 2) for all $0\leq i\leq u-1$, and (2) $\langle {\bf v},{\bf r}^u\rangle=1$ (modulo 2). 
{\em Only} the register cells corresponding to ${\bf v}_j=1$ are being XORed with inputs $A_j$. 
%The expected Hamming weight of $v$ is $k/2$, which is the number of XORs of register cells with the $A_j$ as depicted in \figurename~\ref{fig-htd}. 

Since the $A_j$ are all XORed together in the specified  logical functionality to produce the sum $\sum_j A_j$, the Trojan changes this sum to
$$\sum_j A_j \oplus \sum_{j: {\bf v}_j=1} {\bf r}^i_j = \sum_j A_j \oplus \langle {\bf v},{\bf r}^i\rangle.$$
I.e., the sum remains unchanged until the $u$-th clock cycle when it is maliciously inverted.

The trojan uses an LFSR to generate register values ${\bf r}^i\in \{0,1\}^k$ for each clock cycle $i$ and we assume in our analysis  that all vectors ${\bf r}^i$ behave like random vectors from a uniform distribution. 
Then, it is unlikely that $u$ is more than a small constant larger than $k$ (since every new vector ${\bf r}^i$ has at least probability $1/2$ to increase the dimension by one). 
Therefore, $u\approx k$, hence, the register size of the trojan is comparable to the number of clock cycles before the trojan is triggered to deliver its malicious payload. 
This makes the trojan somewhat contrived (since it can possibly be detected by its suspiciously large area overhead). 

Since inputs $A_j$ can take on any values, any trigger signal $Trig$ must represent a subset of the LFSR register content. Suppose $t(\{Trig\})=j$. Then $Trig$ must represent a subset of ${\bf r}^{u-j}$.
We will proceed with showing a lower bound on $d(\{Trig\})$.
% for 
%the $k$-\textit{XOR-LFSR} trojan. 
%\todo[inline]{We have not defined $L_t(User)$ or $L_0$ yet. Following argument needs to be adapted accordingly.}
%(2) %By our definition of $L_0$, all vectors $r^i$, $0\leq i\leq u-1$, are part of states in $L_0$. 
Consider a projection $P$ to a subset of $d$ register cells; by ${\bf r}|P$ we denote the projection of ${\bf r}$ under $P$, and we call $P$ $d$-dimensional. If ${\bf r}^{u-j}|P \in \{{\bf r}^i|P:0\leq i<u-j\}$, then the wire combination of the  $d$ wires corresponding to ${\bf r}^{u-j}|P$ {\em cannot} represent $Trig$ (otherwise $t(\{Trig\})>j$): if this is the case for all $d$ dimensional $P$, then $Trig$ cannot represent a subset of ${\bf r}^{u-j}$. The probability that  ${\bf r}^{u-j}|P \in \{{\bf r}^i|P:0\leq i<u-j\}$ is at least equal to the probability that $\{{\bf r}^i|P : 0\leq i<u-j\} =\{0,1\}^d$, which is (by the union bound) 
$$\geq 1- \sum_{w\in \{0,1\}^d} Prob(\{{\bf r}^i|P : 0\leq i<u-j\} \subseteq \{0,1\}^d\setminus \{w\})$$
\vspace{-6pt}
$$= 1-\sum_{w\in \{0,1\}^d} (1-1/2^d)^{u-j} \approx 1-2^d e^{-(u-j)/2^d}$$
%\begin{eqnarray*}
%& \geq & 1- \sum_{w\in \{0,1\}^d} Prob(\{{\bf r}^i|P : 0\leq i<u-j\} \subseteq \{0,1\}^d\setminus \{w\}) \\
%& = & 1-\sum_{w\in \{0,1\}^d} (1-1/2^d)^{u-j} \approx 1-2^d e^{-(u-j)/2^d} 
%\end{eqnarray*}
Since there are $\binom{k}{d}\leq k^d/d!$ projections, $Trig$ cannot represent a subset of ${\bf r}^{u-j}$ with probability 
%(taken over all random ${\bf r}^i$)
\begin{equation} \geq (1-2^de^{-(u-j)/2^d})^{k^d/d!} \label{eqLB} \end{equation}
For $d \leq \log (u-j) - \log (\log (u-j) \log k +\log \log k)$,  this lower bound is about $\geq 1/e$.
Since $u\approx k$ and after neglecting the term $\log \log k$, this shows an approximate lower bound on $d(\{Trig\})$, i.e.,
$$\geq \log (k-t(\{Trig\})) - \log (\log (k-t(\{Trig\})) \log k) $$

\begin{figure}[!t]
\centering
\subfloat[$k=256$]{\includegraphics[width=1.7in]{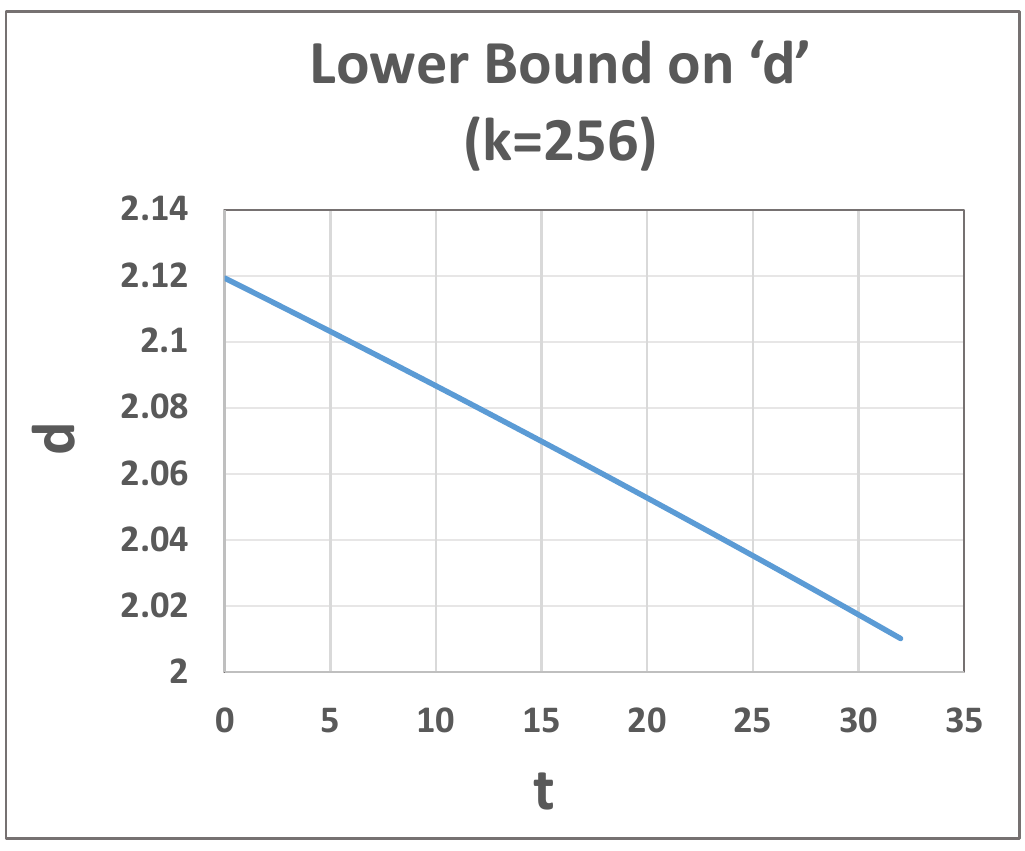}
\label{fig:d_bound_t}}
\hfil
\subfloat[$t=0$]{\includegraphics[width=1.7in]{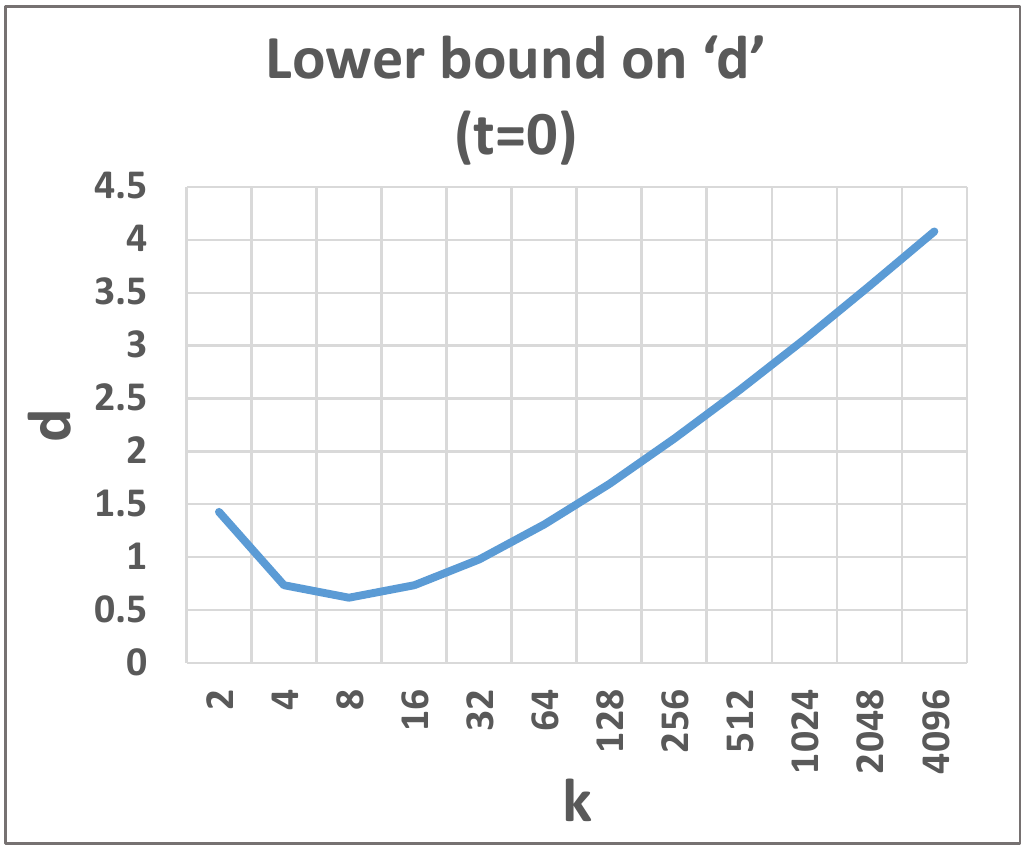}
\label{fig:d_bound_k}}
\caption{Lower bounds on $d$ for $k$-\textit{XOR-LFSR}.}
\label{fig:d_bound_XORLFSR}
\end{figure}

This characterizes the stealthiness of the $k$-\textit{XOR-LFSR}.
%We prove (cf. section \ref{app-htd}) that $k$-\textit{XOR-LFSR} has
%$$ d(\{Trig\}) \geq \log \big(k-t(\{Trig\})\big) - \log \big(\log (k-t(\{Trig\})) \log k \big) $$
%for any trigger signal $Trig$.
In other words, the stealthiness of $k$-\textit{XOR-LFSR} can be increased with $k$ within the acceptable area overhead limits.
\figurename~\ref{fig:d_bound_XORLFSR} shows the lower bounds on $d$ for this HT for fixed values of $k$ and $t$; if $t(\{Trig\}))=0$, then $d(\{Trig\}) \geq \log k -2\log\log k$.
The HT can be designed for any point in the region above the lower bound.
For the HT shown in \figurename~\ref{fig-htd}, clearly %$t(\{Trig\}))=0$ and 
$\alpha(Trig)=0$ since it \textit{always} produces incorrect output immediately once the HT gets triggered.

%% file: sections/evaluation.tex
\section{Characterizing TrustHub Benchmarks}\label{sec:evaluation}

Trusthub~\cite{trust_hub} benchmark suite includes a wide variety of Hardware Trojans including trigger-activated and always-active Trojans; Trojans that use standard IO channels as well as side channels.
In this section, we analyze the TrustHub benchmarks with respect to our definitional framework.
As mentioned earlier, our $H_D$ class of HTs covers majority of the relevant benchmarks from TrushHub. 
Here we present the values of design parameters ($d$, $t$, $\alpha$) for these benchmarks which provide a deeper insight about the stealthiness of these well known HTs.

\begin{table}[t]
\renewcommand{\arraystretch}{1.2}
\setlength\tabcolsep{2pt}
\centering
\caption{Classification of Trusthub Benchmarks w.r.t. our definitional framework}
\label{table:trusthub_table}
\begin{tabular}{|c|c|c|c|c||m{0.7\columnwidth}|}\hline

\multicolumn{2}{|c|}{\multirow{1}{*}{\textit{\textbf{Type}}}}			&
\multicolumn{1}{c|}{\multirow{1}{*}{$d$}}							&
\multicolumn{1}{c|}{\multirow{1}{*}{$t$}}							&
\multicolumn{1}{c|}{\multirow{1}{*}{$\alpha$}}						&
%\multicolumn{1}{c||}{\multirow{1}{*}{$d^*$}}						&
\multicolumn{1}{c|}{\multirow{1}{*}{\textit{\textbf{Benchmarks}}}}	\\  \hline %\cline{1-3} %\hline

\multicolumn{1}{|c|}{\multirow{15}{*}{\textit{\textbf{St}}}}	&
\multicolumn{1}{c|}{\multirow{14}{*}{\textit{\textbf{D}}}}		&
\multicolumn{1}{c|}{\multirow{14}{*}{1}}		&
\multicolumn{1}{c|}{\multirow{4}{*}{0}}						&
       $1/2^{32}$		& BasicRSA-T\{100, 300\} \\ \cline{5-6}
       
&&&& 0.5 	& 	s15850-T100, s38584-T\{200, 300\} \\ \cline{5-6}
&&&& 0-0.25	&	wb\_conmax-T\{100, 200, 300\} \\ \cline{5-6}
&&&& 0-0.87	&	RS232-T\{100, 800, 1000, 1100, 1200, 1300, 1400, 1500, 1600, 1700, 1900, 2000\} \\ \cline{4-6}

&&& \multicolumn{1}{c|}{\multirow{3}{*}{1}}		&
	0.5	&	b15-T\{300,400\}								\\ \cline{5-6}
&&&& 0.5-0.75  &	s35932-T\{100, 200\}						\\ \cline{5-6}
&&&& 0-0.06	  &	RS232-T\{400, 500, 600, 700, 900, 901\} \\ \cline{4-6}

&&& \multicolumn{1}{c|}{\multirow{2}{*}{2}}		&
	0.5		&	vga-lcd-T100, b15-T\{100, 200\}				\\ \cline{5-6}
&&&& 0.87	&	 s38584-T100 							\\ \cline{4-6}

&&& \multicolumn{1}{c|}{\multirow{2}{*}{3}}		&
	$1/2^{32}$	&	BasicRSA-T\{200, 400\}				\\ \cline{5-6}
&&&& 0.5			&	s38417-T100 						\\ \cline{4-6}

&&& 5	&	0.99&	 s38417-T200  \\ \cline{4-6}
&&& 7	&	0.5	&	RS232-T300 \\ \cline{4-6}
&&& 8 	&	0.5	&	s35932-T300 \\ \cline{2-6}

%-----------------------------------------------------------------------------------------------------------------
												&
\multicolumn{1}{c|}{\textit{\textbf{ND}}}		&
\multicolumn{3}{c||}{ N/A }						&
MC8051-T\{200, 300, 400, 500, 600, 700, 800\}, PIC16F84-T\{100, 200, 300, 400\} \\ \hline

%-----------------------------------------------------------------------------------------------------------------
\multicolumn{2}{|c|}{\textit{\textbf{Si}}} 		& 
\multicolumn{3}{c||}{ N/A }						&
AES-T$\lbrace$400, 600, 700, 800, 900, 1000, 1100, 1200, 1300, 1400, 1500, 1600, 1700, 2000, 2100$\rbrace$, s38417-T300, AES-T$\lbrace$100, 200, 300$\rbrace$ \\ \hline 

\end{tabular}
%\vspace{-8pt}
\end{table}

%
%In order to provide concrete guarantees about which trojans can be detected by HaTCh under all circumstances and which ones can be detected under certain conditions, 
\tablename~\ref{table:trusthub_table} shows the relevant\footnote{Not all of the benchmarks from TrustHub are listed in \tablename~\ref{table:trusthub_table}, because some of them have no payload, such as RS232-T200. Similarly the payloads of some other benchmarks are harmless which would be removed by synthesis tools. E.g. RS232-T1800, which just adds three inverters to waste energy.} benchmarks from Trusthub categorized according to our framework.
\textit{St-D} group (i.e. $H_D$ Trojans) is further subdivided based on the properties ($d$, $t$, $\alpha$).
All these Trojans happen to be represented by a single trigger of dimension $d=1$ (i.e., the trigger is a single wire); whereas their corresponding $t$ and $\alpha$ values\footnote{$\alpha$ values show estimated upper bounds on probabilities.} are listed in \tablename~\ref{table:trusthub_table}.

To determine $t$ values of these trojans, we simply count the minimum number of registers between the trigger signal wires(s) and the output port of the IP core.
Since precisely calculating $\alpha$ can be almost impossible for large circuits, we estimate $\alpha$ values through experiments.
We argue that these estimated values closely represent the corresponding actual values of $\alpha$, hence \tablename~\ref{table:trusthub_table} fits the definition of $\alpha$.
In order to estimate $\alpha$ values, we first find the smallest chain of logic gates starting from the trigger signal wire(s) till the output port of the IP core (ignoring any registers in the path).
Then for each individual logic gate, we compute the probability of propagating a logic 1 (considering that the trigger wire(s) get a logic 1 upon a trigger event), e.g. an AND gate has the probability $1/4$ of propagating a logic 1, whereas an XOR gate has the probability $1/2$.
Finally we compute an aggregate probability of propagation by multiplying all the probabilities of each logic gate in the chain, which gives the value $1-\alpha$.
%This number provides a direct estimate of $\alpha$, since a higher number of logic levels means lower probability that the effect of the trigger signal activation will be seen at the output; hence it results in a higher $\alpha$ value.
%As explained earlier, the stealthiness factor $\alpha$ can be reduced by increasing $d$.
%In column $d^*$, \tablename~\ref{table:trusthub_table} shows the required minimum values of $d$ to have an estimated $\alpha=0$.
%It is worth mentioning that HaTCh is able to achieve $\alpha=0$ for these trojans by using locality of only $1$, i.e. $l=1$.
%HaTCh is able to detect all these \textit{St-D} trojans using $d=1$.

Notice that all these \textit{St-D} Trojans have a very low value of $d$ (particularly $d=1$) which reflects their low stealthiness, and hence the fact that these publicly available benchmarks represent only a small subset consisting of simple Trojans.
Even though some of these benchmarks have high values of $\alpha$ (e.g. s38584-T100 and s38417-T200), however in practice, having a very high value of $\alpha$ may not always be useful for the adversary.
Ideally, on one hand, the adversary wants the Trojan to be triggered in the logic testing phase only once, and remain undetected (i.e. by having high $\alpha$) so that the Trojan trigger is whitelisted.
On the other hand, after the testing phase, he wants the Trojan to deliver the payload by disrupting the normal output (i.e. by having low $\alpha$), otherwise the Trojan is not useful for him.
Therefore, the adversary would like to have a sweet spot between the high and low ends of $\alpha$ values.
%This essentially increases the chances for HaTCh to detect the trojan, i.e. either it gets detected (if triggered) in the learning phase (because $\alpha \ll 1$), or it gets detected by the tagging circuitry later.

%Table~\ref{table:trusthub_table} further subgroups \textit{St-D} trojans based on their payload propagation delay $t$, and it can be seen that $t$ is generally pretty small.
%All these trojans happen to have a trigger signal dimension $d=1$.
%Although HaTCh mainly focuses on trigger activated \textit{St-D} trojans, however it can also detect always-active \textit{St-D} trojans since such trojans are meant to always exhibit malicious behavior over the standard I/O channels which can be easily detected by functional testing during the learning phase.
%Notice, however, that always-active \textit{St-D} trojans are practically not useful for an adversary as they have no stealthiness.

%\subsubsection{ST-ND and Si Trojans}
\textit{St-ND}  (i.e. $H_{ND}$ group) and \textit{Si} Trojans are out of scope of this paper.
In our model, some TrustHub benchmarks, e.g. MC8051 series, are considered to be in $H_{ND}$ group because of their flexible design specifications.
The specification of a processor is relatively flexible about the timing/ordering of outputs (e.g. instructions execution) due to some unpredictable factors like interrupt requests etc. 
This flexibility, however, makes it harder for logic testing based tools to verify the functional correctness of the design. 
However, if there exists a precise and strict model for these cores, such HTs can still be analyzed in our model.
%Notice that trigger activated \textit{Si} trojans can actually be detected by HaTCh provided that these trojans do not get triggered during the HaTCh learning phase (so that their trigger related wires still remain blacklisted).
%\textit{Si} trojans exploit side channels to deliver the malicious payload and do not affect the normal functionality over standard I/O channels.
%If such a trojan gets activated during the learning phase, since HaTCh tool will not observe any malicious behavior on the standard I/O channels therefore it would whitelist the trigger related wires.
%On the other hand, if it does not get activated during the learning phase, only the non-malicious wires would be whitelisted and the trigger related wires would still remain in the blacklist leading to trojan detection.
%The same reasoning is applicable to \textit{TA-St-D-NF} trojans.
%Always-active \textit{Si} trojans, however, are out of the scope of HaTCh tool and cannot be detected.
%Since these trojans are not trigger-based, therefore they are always active during the learning phase.
%However such trojans always show malicious behavior in a way that does not harm the normal functionality of the IP core and hence HaTCh considers the circuitry as `normal' and whitelists it.

%% file: sections/related_work.tex
\vspace{6pt}
\section{Related Work}\label{sec:related_work}

Hardware trojans have recently gained significant interest in the security community~\cite{taxonomy},~\cite{usenix08},~\cite{huntforkill}. 
The works~\cite{usenix08} and~\cite{huntforkill} showed how malicious entities can exist in hardware, while Skorobogatov \emph{et al.}~\cite{military} showed evidence of such backdoors in military grade devices.
Nefarious designs have also been deployed and detected in wireless communications devices~\cite{wireless}.
Recent works have mostly focused on detection~\cite{iccd_12} and identification schemes~\cite{iccad_12}, which assess to what extent the pieces of hardware may be vulnerable, and how related trojans can be classified.
State of the art HT detection schemes include UCI~\cite{sp2010}, VeriTrust~\cite{veritrust}, FANCI~\cite{fanci} and DeTrust~\cite{detrust}.
Typically, HT detection techniques only show their detection capabilities for HTs from the TrustHub benchmarks suite, in which all trojans are explicitly triggered.
This explicitness forgoes the lack of implicitness, due to which all the above schemes are able to detect the benchmarked trojans.
These schemes, however, do not cater for higher dimensional ($d>1$) trojans or the added stealthiness because of the implicit malicious behaviors (i.e. $\alpha$).
DeTrust presents a trojan example which bypasses other existing countermeasures, and interestingly it happens to have the dimension $d=2$.
However, this property has not been noticed or analyzed in that paper.
We fill these gaps by providing a detailed and rigorous framework to reason about HT characteristics and their impact on the detection schemes.
%% *************************************************************** 

%\vspace{6pt}
%\textbf{Side-Channel HT Detection Techniques:} 
Further works construct and detect HTs that use side channels~\cite{sidechannel1, sidechannel2, sidechannel3}.
Such HTs remain implicitly on, and have usually no effect on the normal functionality of the circuit.
Side channels include power based channels~\cite{power2}, as well as heat based channels~\cite{power4}.
Power based HTs force the circuit to dissipate more and more power to either damage the circuit, or simply waste energy.
Heat based HTs leak important information via heat maps~\cite{heat}, where highs and lows in heat dissipation can be interpreted as logic 1's and 0's.
%The above works seem to be complete in the sense that the threat model contains side channel vulnerabilities, however, this paper uses threat models that cater for direct contacts between the user and the design.
Our work only focuses on HTs that perturb standard I/O channels; analyzing side channel models/frameworks is out of scope of this paper.
%The presence of a non-zero false negative rate in an adversarial model that allows side channel HTs implies a constant rate of privacy leakage.%\footnote{Therefore, as a design principle, IP cores that need to protect against (maliciously engineered) leakage over side channels must implement leakage resilient key renewal, e.g., as described in \cite{hazay2013leakage}.
%It is outside the scope of this paper to analyze side-channel models/frameworks in existing literature that may lead to tools that can detect side channel HTs with small false negative rates or obfuscate (by adding extra circuitry) the effect of such HTs leading to reduced privacy leakage rates.

%% file: sections/conclusion.tex
\section{Conclusion}\label{sec:conclusion}
We provide the first rigorous framework of Hardware Trojans within which ``Deterministic Trojans'', the class $H_D$ is introduced.
We discover several stealthiness parameters of the Hardware Trojans from $H_D$ which lead us to design a much more stealthy \textit{XOR-LFSR} Hardware Trojan.
This shows that the current publicly known Hardware Trojans are the simplest ones in terms of stealthiness, and hence they represent just the tip of the iceberg at the huge landscape of Hardware Trojans.

We conclude that our framework allows the Hardware Trojan research community to rigorously reason about the stealthiness of different Hardware Trojans and the effectiveness of existing countermeasures.
The Hardware Trojan design principles introduced in this paper encourage and assist in designing new and even stronger countermeasures for highly stealthy and sophisticated Hardware Trojans.